\documentclass[12pt]{iopart}

\newcommand{\uptr}{\hbox{\raise0.0ex\hbox{$\bigtriangleup$}}}
\newcommand{\dntr}{\hbox{\raise0.5ex\hbox{$\bigtriangledown$}}}
\newcommand{\gae}
{\,\hbox{\lower0.5ex\hbox{$\sim$}\llap{\raise0.5ex\hbox{$>$}}}\,}
\newcommand{\lae}
{\,\hbox{\lower0.5ex\hbox{$\sim$}\llap{\raise0.5ex\hbox{$<$}}}\,}
\usepackage{iopams}
\usepackage{graphicx}
\begin{document}

\title{ Scaling in the vicinity of the four-state Potts fixed point}

\author{H.~W.~J. Bl\"ote$^1$, W.-A. Guo$^2$  and M.~P. Nightingale$^3$}

\address{$^{1}$Instituut Lorentz, Leiden University, P.O. Box 9506,
2300 RA Leiden, The Netherlands\\
$^{2}$Physics Department, Beijing Normal University,
Beijing 100875, P.~R.~China\\
$^3$Department of Physics, University of Rhode Island,
Kingston RI 02881, USA}
\ead{$^1$henk@lorentz.leidenuniv.nl, $^2$waguo@bnu.edu.cn,
$^3$nightingale@uri.edu}
\begin{abstract}
We study a self-dual generalization of the Baxter-Wu model,
employing results obtained by transfer matrix calculations of the
magnetic scaling dimension and the free energy.  While the pure
critical Baxter-Wu model displays the critical behavior of the four-state
Potts fixed point in two dimensions, in the sense that logarithmic
corrections are absent, the introduction of different couplings in the
up- and down triangles moves the model away from this fixed point, so that
logarithmic corrections appear.  Real couplings move the model into
the first-order range, away from the behavior displayed by the
nearest-neighbor, four-state Potts model. We also use complex couplings, 
which bring the model in the opposite direction characterized by the
same type of logarithmic corrections as present in the four-state Potts
model.  Our finite-size analysis confirms in detail the existing 
renormalization theory describing the immediate vicinity of the
four-state Potts fixed point.
\end{abstract}

\submitto{\JPA}
\maketitle

\section {Introduction}
\label{sec1}

Once upon a time, the idea originated \cite{Wegner} that marginally
irrelevant scaling fields can introduce logarithmic corrections to scaling.
The two-dimensional, four-state Potts model, for which some exact analysis
is possible \cite{BaxP}, is a celebrated example of this kind of behavior.
Cardy \cite{Car} investigated the presence of such corrections to
the finite-size scaling behavior of this model. If we denote by $v$
the marginal field, and by $L$ the linear dimension of the system,
Cardy considered the effects of multiplicative logarithmic correction
factors of the form $c_k v^k/[1-v\ln(L)/\pi]^k$. His approach
assumed that $v<0$ and $\ln L \gg |v|^{-1}$, so that the $v$-dependence
vanishes to lowest order of $1/\ln L$. The numerical data
upon which Cardy's based his analysis were obtained from
transfer matrix computations for strips of a limited range of widths
$L$~\cite{BN82}, and his analysis was predicated on the
assumption that $|v|$ was sufficiently large.

In this paper we study a generalized Baxter-Wu model in which that
condition is not always satisfied. As a matter of fact, in our case
$|v|$ can be made arbitrarily small,  as discussed in greater detail in
Sec.~\ref{alltheory}. Our purpose is to verify the precise form of
these logarithmic correction factors, without expanding in $1/\ln L$. 

Other ways to vary the marginal field are the introduction of
vacancies  \cite{NBRS,QDB}, of four-spin interactions \cite{SAB}, and
of interactions of a sufficiently long range \cite{QDLGB}.
The marginal field for a $q=4$ Potts model with
16 interacting neighbors was found to be very small \cite{QDLGB}.
However, our analysis will require high-precision data generated by a
transfer-matrix method, which would be restricted to small system
sizes for the latter model.

Sec.~ \ref{alltheory} also provides a summary of the scaling theory near
the four-state Potts fixed point. It also provides a short description
of the generalized Baxter-Wu model, and of the transfer-matrix method
employed to generate the finite-size data.

The scaled magnetic gaps are analyzed in Sec.~\ref{numanscg}, and the
free energy data in Sec.~\ref{numanfre}. We conclude this paper with
a short discussion in Sec.~\ref{disc}.

\section {Theoretical results}
\label{alltheory}
\subsection {Renormalization description of the $q=4$ Potts model}
\label{theory}
The relevant theory is based on the renormalization flow scenario as 
proposed by Nienhuis et al.~\cite{NBRS} for the two-dimensional Potts
model. For the $q=4$ Potts universality class, one considers three
scaling fields, namely the temperature field $t$, the magnetic
field $h$, and the marginal dilution field $v$. The critical range
of the $q=4$ Potts model corresponds to $v<0$.
For $q=4$, the renormalization equations of Nauenberg and
Scalapino \cite{NS} and Cardy \cite{CNS,Car}, and of Salas and
Sokal \cite{SS}, reduce to
\begin{equation}
\hspace*{-26mm}
   \frac{dv(x)}{dx}= \pi b_v v^2(x)\,
,~~\frac{dt(x)}{dx}=\left[\frac{ 3}{2}+2\pi b_t v(x)\right]t(x)\,
,~~\frac{dh(x)}{dx}=\left[\frac{15}{8}+2\pi b_h v(x)\right]h(x)\, ,
\end{equation}
where $x$ parametrizes the scale reduction factor $b$ as $e^x=b$.
The coefficients $b_v=4/\sqrt{3}, b_t=\sqrt{3}/2, b_h=\sqrt{3}/24$
are universal if the field coupled to $v$ is normalized so that its
two-point correlation is $r^{-4}$ \cite{Car}. If $v$ is not normalized,
the ratios $b_h/b_v=1/32$ and $b_t/b_v=3/8$ are still universal.
In the presence of a finite-size parameter $L$, this renormalization
approach leads, for normalized $v$ and a rescaling factor
$L$, to the following scaling equations for $t$, $h$ and $v$:
\begin{equation}
t \to t'=L^{y_t}z^{3/4} t\,  ,~~~~  h \to h'=L^{y_h}z^{1/16} h\, ,~~~~
v \to v'=z v \, ,
\label{vs0}
\end{equation}
where  the scaling factor $z(L)$ of $v$ depends on $L$ as
\begin{equation}
z(L) = 1/[1-\frac{4 \pi}{\sqrt{3}} v\ln L] \, .
\label{vs1}
\end{equation}
 The consequences of these equations are as follows. For a model with
a negative marginal field, $|v|$  will decrease under  renormalization,
but with an anomalously slow rate if $v$ is small. This leads to the
logarithmic corrections seen in the $q=4$ Potts model. But for $v>0$ the
marginal field grows under renormalization, at an increasing rate when 
$v$ becomes larger. Then, the model renormalizes towards a discontinuity
fixed point \cite{NN} and the transition becomes discontinuous,
as seen in the dilute Potts model \cite{NBRS,QDB}, in a Potts model
with four-spin interactions \cite{SAB}, and in models with interactions
of sufficiently long range \cite{QDLGB}.

Thus the singular part $f_s$ of the free energy density $f=\ln Z$ scales as
\begin{equation}
f_s(t,h,v,L) =L^{-2} f_s(L^{y_t}z^{3/4} t,L^{y_h}z^{1/16} h, zv,1) \, ,
\label{fs1}
\end{equation}
Here we consider the case of a model wrapped on an infinitely long
cylinder with circumference $L$.  According to Cardy \cite{Car}, 
for this geometry, expansion of $f_s$ in its argument
$zv$ leads to vanishing contributions in first and second order.
Thus, at $t=h=0$,
\begin{equation}
f_s(0,0,v,L)=\frac{\pi}{6 L^{2}}
\left\{1+\frac{c_3v^3}{[1-\frac{4\pi}{\sqrt{3}}v\ln(L)]^3}+\ldots\right\}\, .
\label{fs2}
\end{equation}
Nevertheless, the analytic part of the free energy may still contain
contributions in first- and second order of $v$.
As noted by Cardy \cite{Car}, $c_3=4 \pi^3 b_v$ is universal if $v$ is
normalized.  The ratio $c_3/(b_v \pi)^3=3/4$ is still universal if $v$
is not normalized. Therefore, if $v<0$ and $L$ sufficiently large,
one has $|v\ln L|>>1$ and one may expand in $1/\ln L$. Then, the 
lowest-order logarithmic correction  term $\pi/[8L^2 (\ln L)^3]$ is
independent of $v$, i.e., universal.
However, in the present work, we will also be interested in the
extension of this range to $|v\ln L| \ll 1$.

The scaling behavior of the scaled, inverse magnetic correlation length
$X_h(v, L) \equiv L/[2 \pi \xi(L,v)]$ can also be expressed as an 
expansion of the scaling function in the marginal field. In this case 
there exists a contribution that depends to first-order on $v$ \cite{Car}:
\begin{equation}
X_h(v,L)=X_h + \frac{p_1 v}{
1- \frac{4 \pi}{\sqrt{3}}v \ln L 
}+\ldots\, .
\label{fs3}
\end{equation}
where $X_h$ is the magnetic scaling dimension \cite{Cardyxi}. The ratio
$p_1\sqrt{3} /4\pi=2 b_h/b_v=1/16$ is universal \cite{Car}.

\subsection {The generalized Baxter-Wu model}
The proper Baxter-Wu model has been solved exactly \cite{BW}.
It belongs to the universality class of the four-state Potts model in two
dimensions, and has the special property that  the marginal scaling
field $v$ is exactly zero. Thus, apart from irrelevant fields, the
Baxter-Wu model sits precisely at
the $q=4$ Potts fixed point, and the logarithmic factors mentioned in
Sec.~\ref{theory} are absent. Here we consider a generalized Baxter-Wu
model with different couplings in the up- and down triangles:
\begin{equation}
{\mathcal H} = -K_{\rm up} \sum_{\uptr} \sigma_i \sigma_j \sigma_k
       -K_{\rm down} \sum_{\dntr} \sigma_l \sigma_m \sigma_n
\label{BW2c}
\end{equation}
where the sums are on the up- and down triangles of the triangular
lattice, and the spins $\sigma$ take the values $\pm 1$ and carry a label 
denoting their lattice site. This model has a line of self-dual
points located at \cite{GHM,genBW}
\begin{equation}
\sinh 2 K_{\rm up} \sinh 2 K_{\rm down} = 1 
\label{BWsd}
\end{equation}
Defining
\begin{equation}
 K_{\rm up} = K+ \Delta K, ~~~~~ K_{\rm down} = K- \Delta K,
\label{rewr}
\end{equation}
the condition for the self-dual line  is rewritten as
\begin{equation}
\sinh 2 K  = \cosh 2 \Delta K 
\label{BWsd1}
\end{equation}
Numerical investigation \cite{genBW} of the model (\ref{BW2c}) has shown
that, for $\Delta K \ne 0$, the model still undergoes a phase transition
at the self-dual point, but its location moves away from the $q=4$
Potts fixed point, in the direction of positive $v$.
Thus the transition becomes discontinuous for $\Delta K\neq0$.

Since $\Delta K$ is a variable parameter, the generalized Baxter-Wu model
enables exploration of the renormalization equations for a range of
values of the marginal field $v$, while the critical point remains exactly
known. In view of the symmetry of the model under interchange of 
$K_{\rm up}$ and $ K_{\rm down}$, one expects that $v \propto \Delta K^2$
to lowest order.  It may thus  seem that this approach is restricted to
the range $v \geq 0$, but we may also include complex couplings
with $\Delta K=\pm{\rm i}\phi$, so that
\begin{equation}
 K_{\rm up} = K+ {\rm i} \phi , ~~~~~ K_{\rm down} = K- {\rm i} \phi \, .
\label{rewi}
\end{equation}
The condition for the self-dual points of the model, Eq.~(\ref{BWsd1}),
then translates into
\begin{equation}
\sinh 2 K = \cos 2 \phi  \, .
\label{BWsd2}
\end{equation}
While individual spin configurations may contribute complex terms to the
partition sum $Z$, the imaginary contributions cancel in systems symmetric
under rotation by $\pi$.
\begin{figure}
\begin{center}
\includegraphics[scale=0.9]{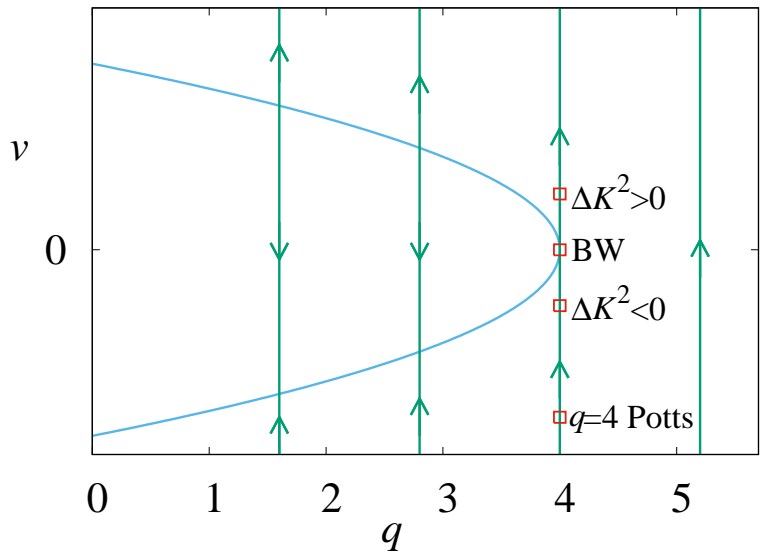}
\end{center}
\caption{The renormalization flow diagram  as proposed by Nienhuis
et al.~\cite{NBRS} in the critical surface of the Potts model,
parametrized by its number of states $q$ and the activity of the
Potts vacancies $v$. The blue curve is a line of fixed points,
consisting of a Potts critical branch (lower part) and a tricritical
branch (upper part).  For $q=4$, $v$ assumes the role of the marginal
scaling field. We thus set $v=0$ at the $q=4$ fixed point. For $v<0$
the scaling field is marginally irrelevant at $q=4$, for $v>0$ it is
marginally relevant. As argued in the text, $v \propto \Delta K^2$ for the
generalized Baxter-Wu model, thus enabling investigation of finite-size
scaling in the presence of a continuously variable marginal field $v$. 
The positions of the $q=4$ Potts model, the Baxter-Wu model (BW) and
the generalized Baxter-Wu model with two arbitrary values of
$\Delta K^2$ are indicated in the figure (red squares).
 }
\label{rgflow}
\end{figure}
Thus we can analyze the scaling behavior of the model for either sign of
the marginal field. In Fig.~\ref{rgflow} we sketch the inferred location
of a few models in this universality class, placed in the wider context
of the surface of phase transitions of the the $q$-state Potts model
and the equivalent random-cluster model \cite{KF}.

\subsection {The transfer matrix}
We employ the transfer-matrix method to calculate the free energies
and the scaled magnetic gaps mentioned in Sec.~\ref{theory}. 
The Ising spins of the Baxter-Wu model are simply coded as binary
digits.
We computed the largest two eigenvalues
by means of a sparse-matrix method \cite{PN79}, which enabled us to handle
the Baxter-Wu model with finite sizes up to $L=27$, with transfer
perpendicular to a set of edges of the triangular lattice. We adopt
the nearest-neighbor distance as the length unit. For real
couplings, we used the algorithm employed in Ref.~\cite{genBW}. 
For the generalized Baxter-Wu model with complex couplings, the transfer
matrix is complex but Hermitian. The eigenvalue spectrum remains
real, and the conjugate-gradient method \cite{CG} could be adapted
to handle this case. 

For transfer parallel to a set of
edges, we obtained finite size data for even system sizes with
$\Delta K=0$ up to $L=30 \sqrt{3/4}$. The transfer matrix for this
case is not symmetric, but it is related to the transpose transfer 
matrix by geometric symmetry operations. We used the tridiagonalization 
method, adapted for this property, to obtain the two leading eigenvalues.

The transfer-matrix method was also applied to the four-state Potts
model on the simple quadratic lattice. We employed the random cluster
representation \cite{KF} combined with the transfer-matrix methods 
described in Ref.~\cite{BN82}. With transfer along a set of edges,
we handled system sizes up to $L=16$, and with transfer along a set of 
face diagonals, system sizes up to  $L=16 \sqrt 2$ (only up to 
$L=15 \sqrt 2$ in the computation of the scaled magnetic gap).

These transfer-matrix algorithms apply to systems that are periodic
with the finite size $L$ in one direction, in the limit of infinite
size for the other direction. 
Let the largest eigenvalue of the transfer matrix that adds one layer
of the system be $\Lambda_0$. Then the free energy density follows as
\begin{equation}
f(L)= \frac {\zeta }{L} \ln \Lambda_0
\end{equation}
where $\zeta$ is the geometrical factor, defined here as the
inverse of the area per site, expressed using the nearest-neighbor
distance as the length unit. It takes the value $\zeta=\sqrt{4/3}$ for
the Baxter-Wu model. The magnetic correlation
length $\xi_{m}(L)$ associated with the second eigenvalue $\Lambda_1$
of the transfer matrix follows as 
\begin{equation}
\xi_{m}^{-1}(L) = \zeta \ln \frac{\Lambda_0 }{| \Lambda_1 |}
\label{xkl}
\end{equation}
This quantity enables the calculation of the scaled magnetic gaps
$X_h(L,v) \equiv L/[2 \pi \xi(L,v)]$.
For conformally invariant models these quantities are equal
\cite{Cardyxi} to $X_h$, but here there are
corrections due to being a nonzero distance away from the
conformally invariant fixed point of the $q=4$ Potts model. 

\section {Numerical analysis of the scaled magnetic gaps}
\label{numanscg}
Finite-size data for $X_h(\Delta K, L)$ were derived for a series
of real and imaginary values of $\Delta K$, using the transfer matrix
with transfer perpendicular to a set of edges, for finite sizes
$L=3,~6,~\ldots,~27$.
Before attempting to fit the numerical data with Eq.~(\ref{fs3}),
we shall check for the possible presence of further corrections to
scaling in Sec.~\ref{preins}. On this basis we shall perform the
actual analysis for the generalized Baxter-Wu model in Sec.~\ref{bwxhfit}.
\subsection {Preliminary inspection  of combined data}
\label{preins}
The numerical results for the scaled gaps $X_h(0, L)$, obtained
by setting $\Delta K=0$, are in a good agreement with the exact
limit $X_h=1/8$. Extrapolation of the finite-size data reproduces
this value with an apparent precision in the order of  $10^{-8}$.
This part of the analysis also included data
obtained by a transfer matrix with transfer along a set of edges as
well as one with transfer in the perpendicular direction.
The corrections to scaling are well described by
\begin{equation}
X_h(0, L)=X_h+\sum_i a_i L^{y_i}
\end{equation}
We found clear signs of corrections with exponents $y_1=-4$ and $y_2=-6$,
and an indication for one with $y_3=-8$. No evidence was seen for other
exponents. The amplitudes of these corrections are different for the two
transfer directions; further we observe that $a_1 \approx 0$ for transfer
along a set of edges.

\begin{figure}
\begin{center}
\includegraphics[scale=1.1]{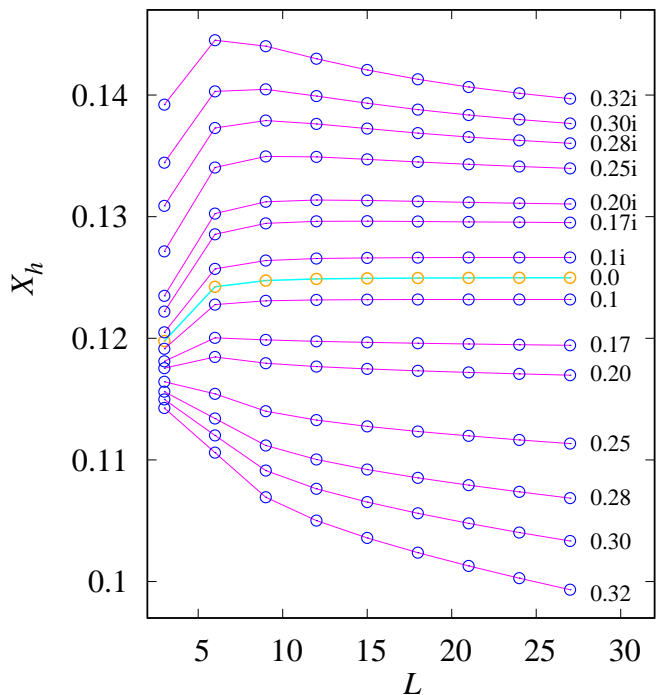}
\end{center}
\caption{Finite-size results for the scaled magnetic gaps $X_h(L)$
of the generalized Baxter-Wu model with several values of $\Delta K$,
as indicated at the right-hand side of the figure. 
 }
\label{xhal1}
\end{figure}

In Fig.~\ref{xhal1} we show the finite-size data for the magnetic
scaled gaps of the critical generalized Baxter-Wu model for several
real and imaginary values of $\Delta K$.  These data are obtained
with the transfer direction perpendicular to a set of edges.
These results reveal a clear picture of the renormalization
flow in the vicinity of the four-state Potts fixed point. For $\Delta K=0$
one observes fast convergence to the exact dimension $X_h=1/8$.
The scaled gaps for imaginary $\Delta K$ display an anomalously slow 
decrease, indeed suggestive of a marginally irrelevant scaling field,
corresponding to $v \propto \Delta K^2 <0$ in Eqs.~(\ref{vs0}),
(\ref{vs1}) and (\ref{fs3}). The results for real values of $\Delta K$ 
display that the renormalization flow, which is still quite weak for
small $\Delta K$, becomes progressively stronger when $\Delta K$ and
$v$ increase, just as expected for a marginally relevant scaling field.

\begin{figure}
\begin{center}
\includegraphics[scale=0.8]{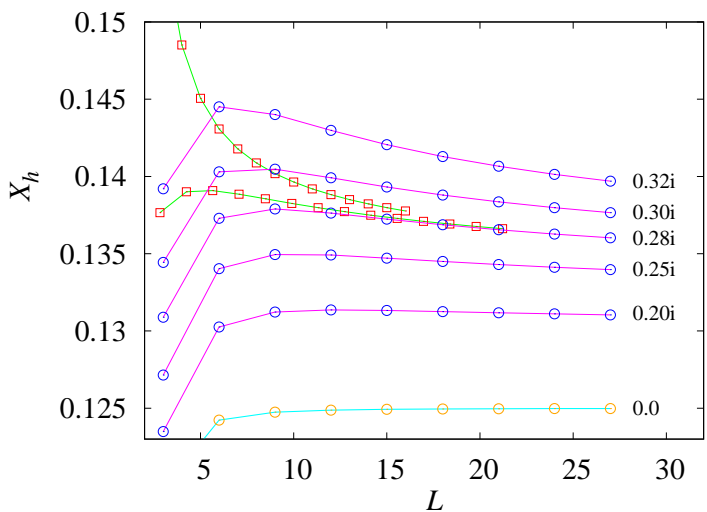}
\end{center}
\caption{Finite-size estimates of the magnetic scaling dimension $X_h$
of the $q=4$ Potts model on the simple quadratic-lattice, and for
the generalized
Baxter-Wu model with several values of $\Delta K$ as indicated at the
right-hand side of the figure. The red squares show data for the $q=4$ Potts
model, with transfer directions along a set of edges (upper squares) and
along a diagonal direction (lower squares). These data apply to finite
sizes (horizontal axis)  up to 16 and $15 \sqrt 2$ respectively.
The data for the generalized Baxter-Wu model are shown as circles, and
are obtained using a transfer matrix perpendicular to a set of edges,
for system sizes up to $L=27$.
 }
\label{xhall}
\end{figure}

\begin{figure}
\begin{center}
\includegraphics[scale=0.6]{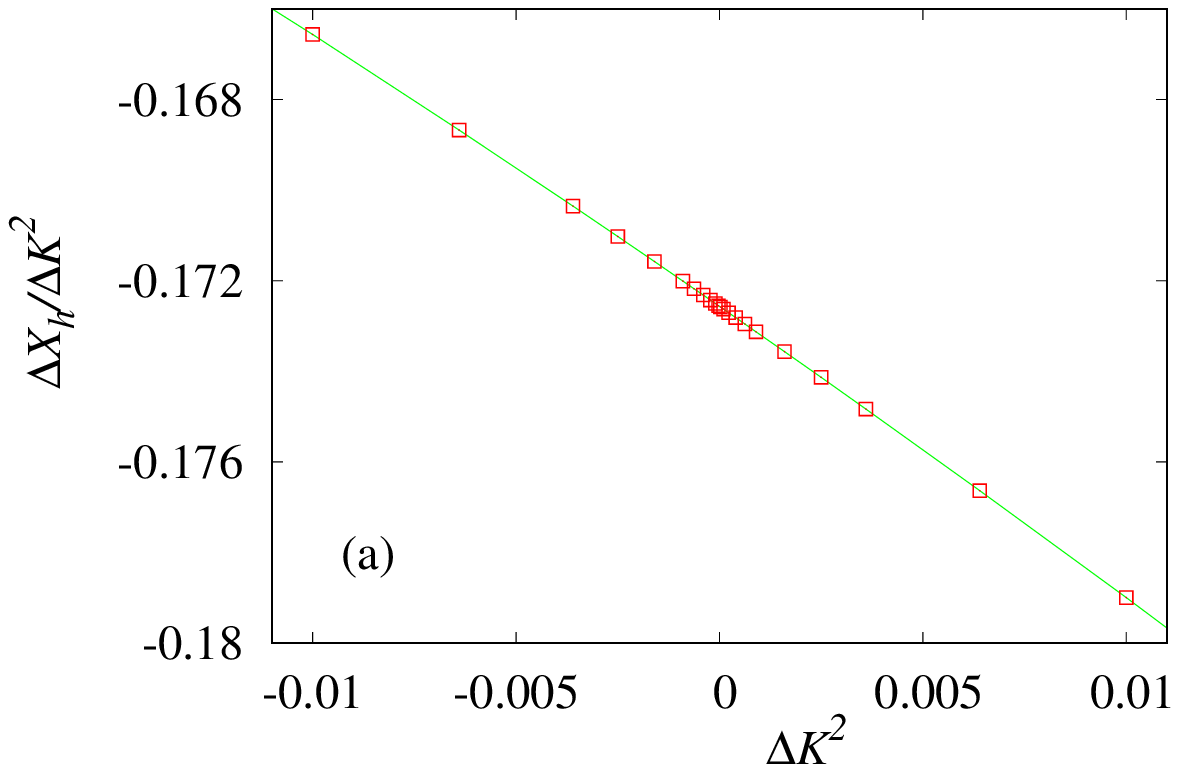}
\includegraphics[scale=0.6]{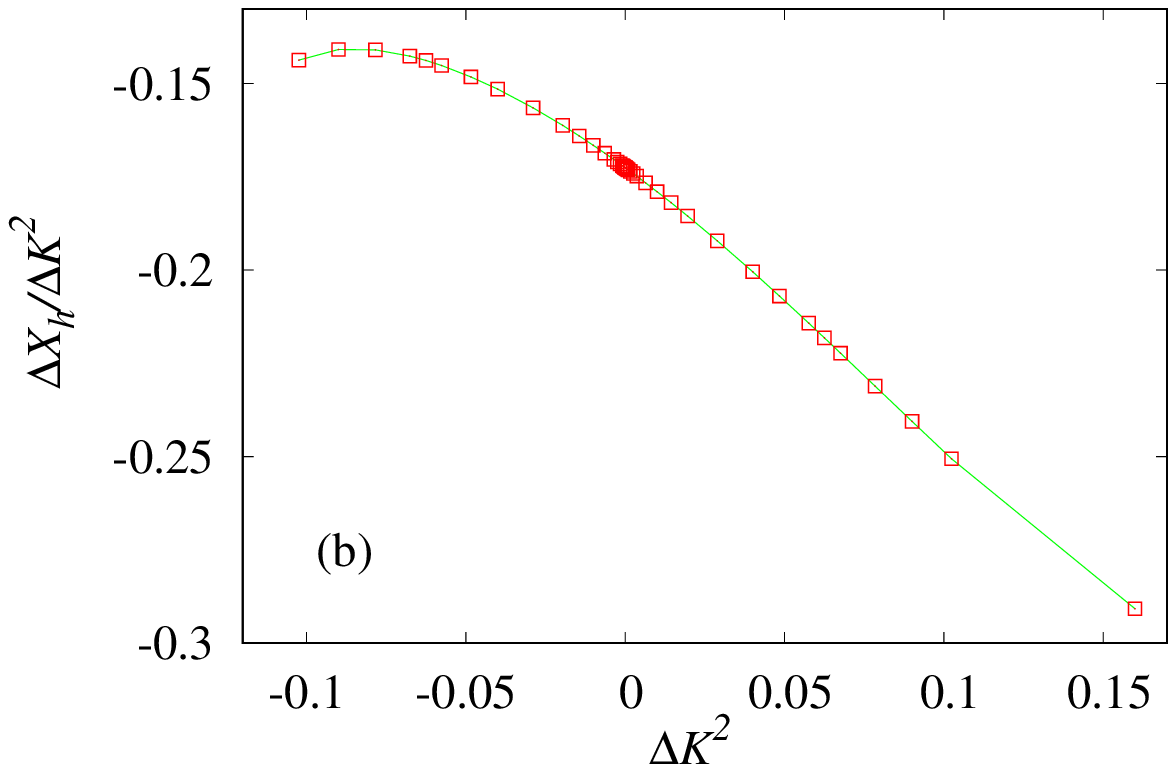}
\end{center}
\caption{Dependence of the scaled gap, expressed by the quantity 
$\Delta X_h/\Delta K^2$ on the marginal field parametrized
by $\Delta K^2$. These data apply to a fixed finite size $L=27$ in
$\Delta X_h=X_h(\Delta K,L)-X_h(0,L)$.
For small marginal fields (a), one finds almost
linear behavior, but on a larger scale (b), nonlinearities become
prominent.}
\label{xh27}
\end{figure}

For comparison, we also show similar results for the four-state Potts
model on the square lattice. Fig.~\ref{xhall} shows the scaled magnetic
gaps for two transfer directions, together with part of the results
for the generalized Baxter-Wu model.
For small $L$, the difference between the two sets of Potts data display
a rapidly decaying $L$-dependence, indicating the presence of terms
with power-law dependence on $L$.  For larger $L$ the dependence
becomes rather weak, similar to the generalized Baxter-Wu results.
These data suggest that the marginal field in the Potts model with
nearest-neighbor interactions on the simple quadratic lattice
corresponds to about $\Delta K=0.28i$ in the generalized Baxter-Wu model.
Our next task is to check to what extent  Eq.~(\ref{fs3}) can describe
these data. At the largest system size $L=27$ and small marginal fields
$v$, one expects that $\Delta X_h \equiv X_h(\Delta K,L)- X_h(0,L)$  
is proportional to $\Delta K^2 \propto v$, with a next order
contribution as $\Delta K^4$ due to the $v$ in the denominator.
Dividing out $\Delta K^2$, the expected linear behavior is reproduced
in Fig.~\ref{xh27}a. For larger marginal fields, higher-order terms
become important. The maximum in Fig.~\ref{xh27}b indicates that we
will require extra terms in addition to the term proportional to
$\Delta K^2$. 
\subsection {Fitting the $X_h$ data for the generalized Baxter-Wu model}
\label{bwxhfit}
After obtaining the finite-size data for $\Delta X_h(\Delta K, L)$, and
thus having suppressed the power-law corrections to scaling that still
occur in the $\Delta K=0$ Baxter-Wu model, we applied least-squares fits
to the quantity $\Delta X_h/\Delta K^2$, based on Eq.~(\ref{fs3}) as follows:
\begin{displaymath}
\hspace*{-20mm}[\Delta X_h(\Delta K, L)]/\Delta K^2 \approx
\sum_k \frac{s_k\Delta K^{2k-2}}{[1-u \ln(L)\Delta K^2]^k} 
+ \sum_l \frac{a_l \Delta K^{2l-2} }{L^{2} [1-u \ln(L)\Delta K^2]^l} + \ldots
\end{displaymath}
\begin{equation}
+ \frac{b_1}{L^2} + \frac{b_2}{L^3} + \ldots \, .
\label{xbwfit}
\end{equation}
The parameter $u$, which remains to be determined, fixes the relation in
lowest order between the marginal field $v$ and $\Delta K^2$ as 
$u \Delta K^2= 4 \pi v/\sqrt{3}$.
\vskip 5mm
\begin{table}[htbp]
\centering
\caption{
Least-squares fits to the finite-size data for $\Delta X_h/\Delta K^2$ of
the generalized Baxter-Wu model with transfer perpendicular to a set of edges.
Estimated uncertainties in the last decimal place are shown between
parentheses, ``f''means that the parameter was fixed.
These error margins are based on statistical analysis and may, at most,
serve as an order-of-magnitude estimate. A better indication of these
errors is provided by the differences between the fits.
The residual $R$, defined in the text, is also listed.}
\vspace{3mm}
\begin{tabular}{||c||c|c|c||}
\hline
param &     fit 1         &      fit 2     &     fit 3       \\
\hline
$a_1$ &$ 0.76   $  (8) & $ 2.14  $   (7) &$ 0.82 $   (7)     \\
$a_2$ &$-5.6    $  (4) & $-12.5$     (3) &$-5.6  $   (3)     \\
$b_1$ &$-1.2    $  (1) & $-2.67$     (5) &$-1.52 $   (7)     \\
$b_2$ &$ 9.53   $  (3) &  10.46      (4) &  14.4     (7)     \\
$b_3$ &$ 0      $  (f) &   0         (f) & -28       (4)     \\
$s_1$ &$-0.172416$ (3) & $-0.172355$ (4) &$-0.17226$ (1)     \\
$s_2$ &  0.934     (5) & 0.892       (4) & 0.931     (4)     \\
$s_3$ &$-5.03   $  (7) & $-4.48$     (5) &$-5.04 $   (5)     \\
$s_4$ &$ 42.4   $  (7) & $ 40.0$     (5) & 44.0      (5)     \\
$s_5$ &$-505    $ (70) & $-474 $    (19) &$-486$    (25)     \\
$  u$ &$-2.74   $  (1) & $-2.68 $    (1) &$-2.740$   (7)     \\
$R/n$&$6.8\times10^{-19}$&$3.9\times 10^{-19}$&$1.7\times 10^{-18}$\\
$L_{\rm min}$& 12      & 15              &12                 \\
$|\Delta K|_{\rm max}$&0.12& 0.14        &0.14               \\
\hline
\end{tabular}
\label{Xhfits}
\end{table} 
The amplitude of the $k=1$ logarithmic term in the first
sum is independent of $\Delta K^2$ and should, for relatively small
$\Delta K^2$, yield the dominant contribution to $\Delta X_h/\Delta K^2$.
The amplitude $s_1$ should thus be easily be resolved, without much
interference from the second-order term which is proportional to
$\Delta K^2$, and without much interference from the denominator. 
Then, the next step is to include larger $|\Delta K^2|$, in the hope that
the fit can disentangle the contribution with amplitude $s_2$ from the
$\Delta K$ dependence of the denominator of the first logarithmic term,
which also behaves as $\Delta K^2$ for not too large $|\Delta K|$. 

In addition to the unknown parameters in the fit formula, we are also
interested in their margin of uncertainty. Since the deviations between
numerical data $\Delta X_h(\Delta K, L)$ and $\Delta_c X_h(\Delta K, L)$
as computed from the fit formula are not of a statistical origin, the
usual procedure is not applicable. We first define the residual  as
\begin{equation}
R=\sum |\Delta K|^2[\Delta X_h(\Delta K, L)-\Delta_c X_h(\Delta K, L)]^2/L^4 
\end{equation}
which takes into account the data with small $|\Delta K|$ can be fitted
very precisely with small contributions to higher-order terms in 
$\Delta K^2$. The additional weight factor $L^{-4}$ accounts for the 
reduced precision of the transfer-matrix results for the largest $L$.
After minimizing $R$, the scale of the deviations between the actual and
computed values of $\Delta X_h$ follows as $L^2 \sqrt(R/n)/|\Delta K|$
where $n$ is the number of data points.  
Next we ask the question what deviations in the fit parameters are still
acceptable. For data with errors of a random nature, the answer is
provided by standard statistical analysis based on numerical errors 
in $\Delta X_h$ of magnitude $L^2 \sqrt(R/n)/|\Delta K|$. In the present
case this answer does not apply, but it may still serve as an order of
magnitude estimate. A more reliable estimate of the margin in the fit
parameters is obtained by varying the number of fitted parameters, the
range of $\Delta K^2$, and the cutoff $L_{\rm min}$ at small system sizes.

We included fits with only one logarithmic term, the one with amplitude
$s_1$. These yielded only consistent results in a rather narrow
range of $\Delta K^2$, so that the parameters $u$ and $s_1$ could not
reliably be determined.  Results of various other fits with more than
one logarithmic term rather consistently yielded $s_1\approx -0.1724$,
and $u \approx -2.74$. For a few of these fits, the parameters are listed
in Table \ref{Xhfits}.
We also investigated the effects of a nonlinear term in $u$ proportional
to $\Delta K^4$. Its influence is very similar to that of adding a next
term in the sum on $k$ in Eq.~(\ref{xbwfit}), and we were unable
to find a significant result for its amplitude. Since the coefficients
$b_i$ may depend on the marginal field, we also included a term
proportional to  $L^{-2} \Delta K^2$, but its amplitude is small and 
did not significantly reduce $R$.

With the exception of $a_3$ and $a_5$, most parameters seem reasonably
well determined, in particular $u$, $s_1$ and $s_2$ describing the
leading logarithmic terms. The third fit covers the widest range
of system sizes and $\Delta K$ values. It yields a universal ratio
$s_1/u=0.0629~(1)$, in a reasonable agreement with Cardy's exact
value \cite{Car} $s_1/u=1/16=0.0625$. Furthermore, these parameters enable
the determination of the marginal field. Using the normalization 
mentioned in Sec.~\ref{alltheory}, a comparison between Eqs.~(\ref{fs3})
and (\ref{xbwfit}) shows that $p_1 v= s_1 \Delta K^2$. With
$p_1=\pi/(4 \sqrt 3)$ \cite{Car}, the numerical result for the
amplitude $s_1$ then determines the marginal field as
$v \approx -0.380 \Delta K^2$.

\subsection {Analysis of the $q=4$ Potts model}
\label{xq4}
In analogy with Sec.~\ref{bwxhfit}, the finite-size data for
$X_h(L)$ were fitted by 
\begin{equation}
X_h(L)=X_h + a_1/L^2 +a_2/L^4 +a_3/L^6 +
\sum_k p_k /(1-w \ln L )^k +\ldots \, .
\end{equation}
The residual $R$ is simply defined as the sum of the squared deviations
between the actual and computed values if $X_h(L)$.
The value of $\chi^2$ increases rapidly when the minimum system size
$L_{\rm min}$ is decreased below 10.
We were unable to find reliable results from fits with all parameters left
free. For this reason, we fixed the known amplitude ratio $p_1/w=1/16$,
and $p_2/w^2=0.124$, $p_3/w^3=0.245$, $p_4/w^4=0.7804$, and $p_5/w^5=3.15$,
as found from the amplitudes for the generalized Baxter-Wu model.
\vskip 5mm
\begin{table}[htbp]
\centering
\caption{\label{q4potx}
Least-squares fits to the finite-size data for $X_h(v,L)$ of
the four-state Potts model for two transfer directions. The second
column specifies the transfer direction, along a set of edges (e) or
diagonals (d). These fits use a small system-size cutoff at
$L_{\rm min}=l_{\rm min}$ for the transfer matrix along a set of edges,
and at $L_{\rm min}=\sqrt 2 l_{\rm min}$ for the diagonal transfer matrix. 
The amplitudes $p_i$ are fixed with respect to the parameter $w$, using
the amplitude ratio as found for the Baxter-Wu model. Although $w$ is the
only adjustable parameter describing the logarithmic correction in this
fit, the residuals per data point, listed as $R/n$, are rather small.
The error margins in the last decimal place, shown between parentheses,
are based on statistical analysis and may, at most, serve as an
order-of-magnitude estimate.}
\vspace{3mm}
\begin{tabular}{|c|c|c|c|}
\hline
param     &transfer&  fit 1    &   fit2         \\
\hline
$a_1$     & e & $ 0.257  $(10) &  0.269   (10)  \\
$a_2$     & e & $ 2.7    $ (2) &  3.0      (2)  \\
$a_3$     & e & $-0.41   $ (3) &  $-0.44$  (3)  \\
$a_1$     & d & $ 0.11   $ (2) &  0.011    (2)  \\
$a_2$     & d & $ 1.7    $ (6) &  1.5      (7)  \\
$a_3$     & d & $-0.37   $ (5) &  $-0.38$  (6)  \\
$p_1$     & - & $ 0.01427$ (f) &  0.01427  (f)  \\
$p_2$     & - & $ 0.00647$ (f) &  0.00646  (f)  \\
$p_3$     & - & $ 0.00292$ (f) &  0.00291  (f)  \\
$p_4$     & - & $ 0.00212$ (f) &  0.00212  (f)  \\
$p_5$     & - & $ 0.00196$ (f) &  0.00195  (f)  \\
$l_{\rm min}$&-&  10           &   11           \\
$w  $     & - & $-0.22826$ (10)&$-0.22818$ (10) \\
$R/n$& - & $7 \times 10^{-15}$ &$1.5\times 10^{-15}$\\
\hline
\end{tabular}
\end{table}
While the numerical uncertainties as listed in Table \ref{q4potx} 
depend on the somewhat arbitrary procedure followed here, and may be
underestimated, a comparison between a number of different fits 
still suggests that the value of $w$ is approximately correct. The
value also determines the marginal field via $4 \pi v/\sqrt 3=w$ as
$v \approx -0.0315$, with the normalization mentioned in Sec.~\ref{alltheory}.
We note that the value $\Delta K \approx 0.28i$, suggested by the 
data in Fig.~\ref{xhall}, for the generalized Baxter-Wu model that has a
marginal field comparable  to that of the $q=4$ Potts model, corresponds
to $w=-0.22$. This is indeed close to the values in Table \ref{q4potx}.

\section {Numerical analysis of the free energy} 
\label{numanfre}
\subsection {Preliminary result for the generalized Baxter-Wu model}
\begin{figure}
\begin{center}
\includegraphics[scale=0.8]{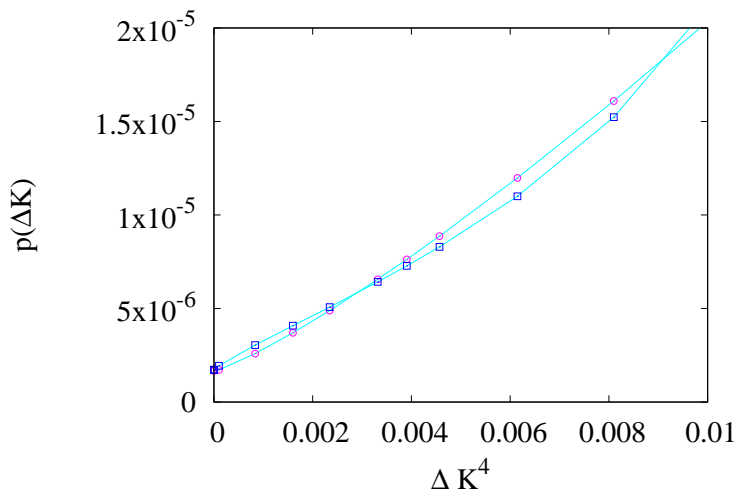}
\end{center}
\caption{Plot of the quantity $p(\Delta K)$ versus $\Delta K^4$.   
It describes the dependence of the singular part of the free energy
of the generalized Baxter-Wu model on the square of the marginal field
which relates to $\Delta K$ as $v \propto \Delta K^2$.
>From Eq.~(\ref{fs2}) one would expect, as long as $v$ is not too large,
a straight line through the origin. 
The data points for real couplings are shown as red circles, those for
complex couplings as blue squares.
}
\label{df4}
\end{figure}

The critical free energy density is exactly known \cite{BWf} for 
$\Delta K=0$ as $ f(\Delta K=0,L=\infty)=\zeta/2 \ln 6$.
The geometric factor $\zeta$ is included because the free energy per 
lattice site is not equal to the free energy density. We can thus
estimate the conformal anomaly $c$ from the finite-size data
$f(\Delta K=0,L)$, from the relation \cite{CDL,BCN,Aff}
\begin{equation}
 f(\Delta K=0,L)- f(\Delta K=0,\infty) \simeq 6       c/(\pi L^2) \, .
\end{equation}
Analysis of the finite-size data for $f(\Delta K=0,L)- f(\Delta K=0,\infty)$
reproduces the exactly known value $c=1$ within a margin of the order of
$10^{-8}$.  
Next we check to what extent Eq.~(\ref{fs2}) describes the $v$ dependence
of the critical free energy. We attempt to get rid of the bulk free energy at
$\Delta K=0$, and of the conformal contribution by defining
\begin{equation}
p(\Delta K)\equiv[f(\Delta K,27)-f(\Delta K,24)-f(0,27)+f(0,24)]/\Delta K^2
\label{pdef}
\end{equation}
Since the finite sizes $L=24$, 27 are presumably large enough, it is 
supposedly justified to interpret $p(\Delta K)$ in terms of the
asymptotic scaling properties of singular part of the free energy.
Since the marginal field $v \propto \Delta K^2$, one may expect,
on the basis of Eq.~(\ref{fs2}), that $p(\Delta K)$ behaves as
$z^3(L) v^2$. Then, for sufficiently small values of
$\Delta K^2$,  $p(\Delta K)$ should be proportional to $\Delta K^4$.
This is supposed to hold for real as well as complex couplings.
In order to check for possible corrections to scaling not included in
Eq.~(\ref{fs2}), the quantity $p(\Delta K)$ is plotted in Fig.~\ref{df4}
versus $\Delta K^4$. The smallest value of $|\Delta K|$ is 0.005, yielding
points at an unresolved distance to the vertical axis in Fig.~\ref{df4}.
One can make the following observations:
\begin{itemize}
\item
In the limit $\Delta K \to 0$, the line does not pass through the origin. 
This can be simply explained
by another $L$-dependent term in the free energy, for instance one
proportional to $v L^{-4}$, and thus also proportional to $\Delta K^2$,
which contributes a constant to $p(\Delta K)$.
\item
The line in the figure consists of two branches, of which one corresponds
to real values of $\Delta K$ (first-order range, $v>0$), and the other
to imaginary $\Delta K$ (continuous transitions, $v<0$). The parabolic
approach of these branches to the vertical axis indicates the presence
of a small $L$-dependent contribution proportional to $v^2$ in $f$.
\item
The two branches display intersections. The curvature leading to
the first intersection is in line with the presence of
a term with amplitude $v$ in the denominator. The second intersection
corresponds to a higher odd power of $v$.
\item
The branches display a general upward curvature  that indicates the
existence of an $L$-dependent contribution in $f$ with a higher even
power of $\Delta K$.
\item
Apart from the above observations, the expected linear dependence on
$\Delta K^4$, as predicted by Eq.~(\ref{fs2}), is still visible to some
extent in this plot for $\Delta K^4 \lae 0.005$. 
\end{itemize}
These findings provide some useful information for a more detailed analysis
of the free energy.

\subsection {Preliminary results for the $q=4$ Potts model}
Finite-size data were calculated for the free energy of the
square-lattice model, using transfer directions along a set of edges
as well as along a diagonal. These transfer matrices add layers
consisting up to 16 spins. 
Since $L$ denotes the strip width in units of lattice edges, the
data from the diagonal transfer matrix apply to finite sizes up to
$L=16 \sqrt{2}$.
\begin{figure}
\begin{center}
\includegraphics[scale=0.7]{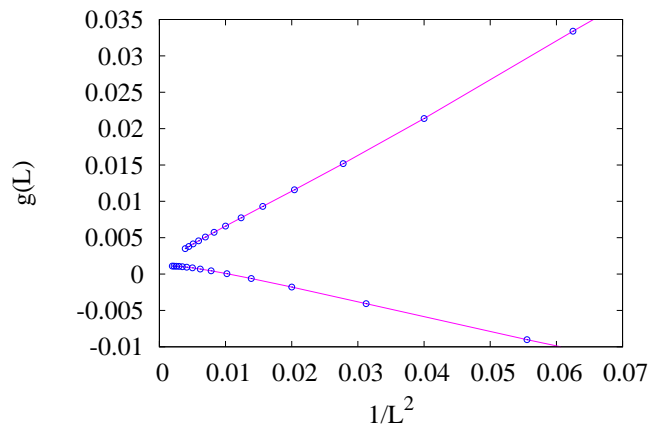}
\end{center}
\caption{Excess critical free energy of the square-lattice $q=4$ Potts
model, shown as the quantity $g(L) \equiv L^2[f(L)-f(\infty)]-\pi/6$
versus $L^{-2}$. The upper set of 
data points is obtained using the transfer matrix for transfer parallel
to a set of lattice edges, the lower set uses transfer in the diagonal
direction. The latter data use a finite size as $\sqrt{2}$ times the
strip width in units of face diagonals.
}
\label{df2}
\end{figure}
In order to find out what contributions to the free energy occur,
in addition to the bulk term $f(\infty)$ and the conformal term 
$\pi/6L^2$, we define the quantity $g(L)$ as
\begin{equation}
g(L) \equiv L^2[f(L)-f(\infty)]-\pi/6 \, .
\label{gdef}
\end{equation}
The data in Fig.~\ref{df2} reveal large contributions to $g(L)$
proportional to $L^{-2}$, corresponding to $L^{-4}$ terms in $f(L)$.
Their amplitudes are seen to depend on the orientation of 
the finite-size direction with respect to the lattice.

The data do not approach the limit $g=0$ at $L=\infty$ linearly,
which is in line with the expected existence of a contribution
with a logarithmic dependence on $L$.

\subsection {Comparison of the generalized Baxter-Wu and the Potts model}
\label{frebwp}
\begin{figure}
\begin{center}
\includegraphics[scale=0.5]{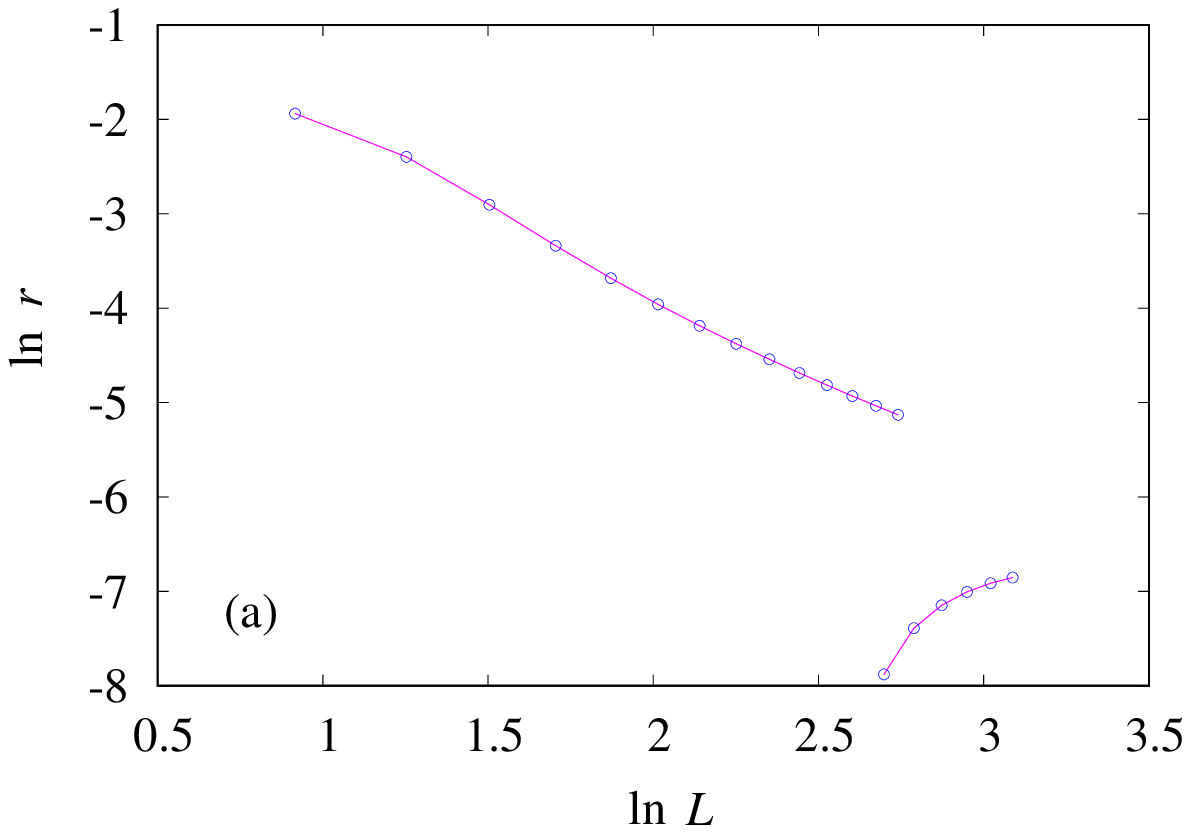}
\includegraphics[scale=0.5]{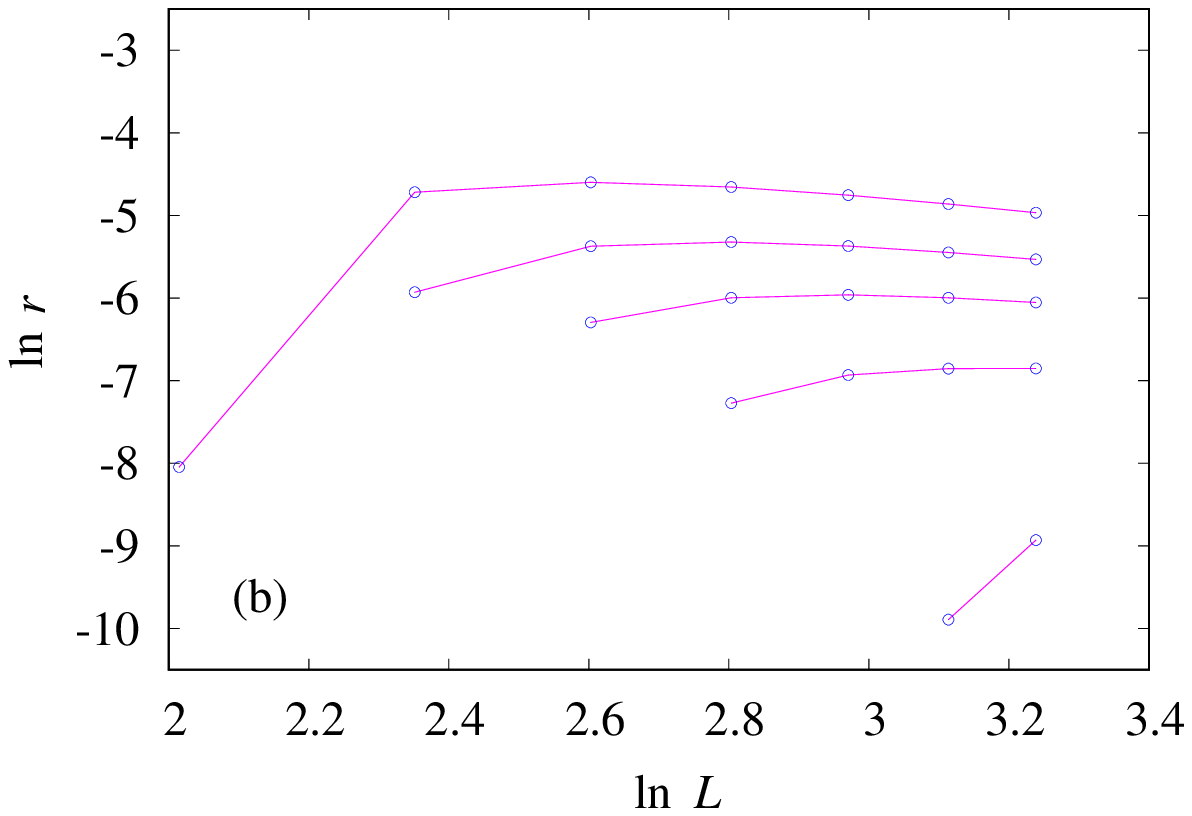}
\end{center}
\caption{Excess critical free energy of the square-lattice $q=4$ Potts
model, as expressed by the quantity $\ln r(L)$, for two transfer
directions (a), and for the generalized Baxter-Wu model (b) with
$\Delta K/i=0.2$, 0.25, 0.28, 0.3 and 0.32. Larger $\Delta K/i$
corresponds to larger $\ln r$ along the vertical scale. The quantity
$r(L)$ changes sign as a function of $L$ for the Potts model with diagonal
transfer, and also for the generalized Baxter-Wu model. The negative
entries occur at small $L$ and are left out of these figures.  }
\label{q4bwd}
\end{figure}

The bulk free energy is, unlike the $q=4$ Potts model, not known for
the generalized Baxter-Wu model. But we can remove it, by taking
differences between consecutive finite sizes.
To this purpose we define
\begin{equation}
r(\overline{L})\approx [f(L_1)-f(L_2)]/(L_1^{-2}-L_2^{-2}) -\pi/6
\label{fcor}
\end{equation}
where $\overline{L}=(L_1+L_2)/2$. This eliminates, at the same time,
the conformal contribution at the Potts fixed point.
Figures \ref{q4bwd} show the numerical results for $r(L)$
for the $q=4$ Potts model, and for the generalized Baxter-Wu model with
$\Delta K/i=0.2$, 0.25, 0.28, 0.3 and 0.32.
The figure suggests that the magnitude of the correction to $f$ due to the
marginal field in the $q=4$ Potts model corresponds roughly to
$\Delta K=0.28i$ in the generalized Baxter-Wu model, consistent with the 
results from the $X_h(L)$ analysis.
\subsection {Fit of the generalized Baxter-Wu free energies}
\label{bwffit}
First we define
\begin{equation}
y(L,\Delta K) \equiv [f(L,\Delta K)-f(L,0)]/\Delta K^2
\label{ydef}
\end{equation}
>From Eq.~(\ref{fs2}), and the observations made in Sec.~(\ref{numanfre})
we expect that
\begin{displaymath}
y(L,\Delta K) = \sum_{j=0,1,\ldots} a_j \Delta K^{2j}+\sum_{k=3,4,\ldots}     
 L^{-2} \frac{t_k\Delta K^{2k-2}}{[1-u \ln(L)\Delta K^2]^k}
\end{displaymath}
\begin{equation}
+ \sum_{l=1,2,\ldots} h_l\Delta K^{2l} /L^4+\ldots\, ,
\label{yfit}
\end{equation}
In the numerical analysis, we found that is not well feasible to clearly
resolve the logarithmic contributions with amplitudes $t_3$, $t_4$, ...
when all fit parameters are left free. For this reason
we fixed the parameter $u=-2.74$ as found in Sec.~\ref{numanscg} and
fitted the amplitudes $t_k$. Since the universal amplitude ratio
$t_3/u^3=\pi/8$ is exactly known \cite{Car,DotFat} we have a
consistency check available.
The fitting procedure involved the minimization of the residual $R$
defined by $R= \sum (y_c-y)^2/\Delta K^4$ where $y_c$ is computed from 
Eq.~(\ref{yfit}). The roughly estimated accuracies are, in analogy
with those in Table \ref{Xhfits}, based on the assumption of randomness.
Also for these results one has therefore to estimate the error margins
independently by applying several variations of the fitting procedure.

The parameter values as found from two typical fits are listed in  
Table \ref{fbwfit}.
\vskip 5mm
\begin{table}[htbp]
\centering
\caption{\label{fbwfit}
Results of least-squares fits to the quantity $y(L)$ describing
the dependence of the generalized Baxter-Wu free energy on $\Delta K^2$ 
which parametrizes the marginal field. Errors following from the
least-squares criterion are shown between parentheses, ``f''means
that the parameter was fixed.}
\vspace{3mm}
\begin{tabular}{|c||c|c|}
\hline
param &  fit 1           &  fit 2    \\
\hline
$a_1$ & 1.3075574 (2) & $ 1.3075568$ (2)      \\
$a_2$ & 0.157499 (12) & $ 0.157493 $ (9)      \\
$a_3$ & $-1.4422$ (5) & $-1.4429   $ (4)      \\
$a_4$ & $ 5.13$   (2) & $ 5.13     $ (1)      \\
$a_5$ & $-21.0$   (2) & $-21.2     $ (1)      \\
$a_6$ &   70     (11) & $ 74       $ (3)     \\
$h_1$ & $-0.61  $ (2) & $-0.57     $ (2)     \\
$h_2$ & $ 10.4  $ (6) & $11.4      $ (5)     \\
$h_3$ &  878     (38) & $ 606    $  (25)     \\
$t_3$ & $-9.5$    (2) & $-8.09   $  (16)     \\
$t_4$ &  67       (3) & $ 54     $   (2)     \\
$ u $ & $-2.74$   (f) & $-2.74   $   (f)      \\
$R/n$ & $0.6\times 10^{-16}$ & $1.0\times 10^{-16}$\\
$|\Delta K|_{\rm max}$&0.2& 0.22                     \\
$L_{\rm min}$& 15     &  15                 \\
\hline
\end{tabular}
\end{table}
The condition $t_3/u^3=\pi/8$ appears to be approximately satisfied,
but there is still an uncertainty margin of the order of ten per cent.

\subsection {Fit of the $q=4$ Potts free energy}
Taking into account the logarithmic correction term in Eq.~(\ref{fs2}),
and a few other terms that were found necessary, we fitted the following 
expression to the data for $g(L)$ defined by Eq.~(\ref{gdef})
\begin{equation}
g(L)=a_1 L^{-2}+a_2 L^{-4}+a_3 L^{-4}/\ln L +
\sum_{i=3,4,\ldots} s_i/(1 -w \ln L)^i+\ldots\, ,
\label{gs2}
\end{equation}
where $a_1$ to $a_3$ were allowed to be different for the two transfer
directions, while the $a_i$ and $w$ were taken to be the same. The term
with $L^{-4}/\ln L$ was suggested by the $\chi^2$ criterion. We do not
expect confusion between these $s_i$ and the $s_k$ in Eq.~(\ref{xbwfit}).

Satisfactory fits could be obtained to free-energy data for finite
sizes of 8 and more lattice edges or face diagonals. The residual is
defined here by $R=\sum [g(L)-g_c(L)]/L^2$, where $g_c(L)$ is the value
found from Eq.~(\ref{gs2}). The residual is used 
again to get a rough idea about the numerical error margins in the 
results for the fitted parameters.
These fits were unable to simultaneously determine the parameters
$w$ and $s_i$, in the sense that the estimated error margins exceeded
the actual values. We thus set $w =-0.2282$ as found in Sec.~\ref{xq4}.
Some typical results for the fitted parameter values are listed in
Table \ref{q4potg}.
\vskip 5mm
\begin{table}[htbp]
\label{q4potg}
\centering
\caption{
Results of least-squares fits to the quantity $g(L)$ describing 
the residual free energy density of the $q=4$ Potts model. The second
column specifies the transfer direction, along a set of edges (e) or 
diagonals (d). }
\vspace{3mm}
\begin{tabular}{|c|c|c|c|}
\hline
param &transfer&  fit1   &  fit 2    \\
\hline
$a_1$ & e & $ 0.3911$  (2)&$0.3918 $  (2) \\
$a_2$ & e & $-2.39  $  (2)&$-2.53  $  (3) \\
$a_3$ & e & $ 5.49  $  (3)&$ 5.70  $  (4) \\
$a_1$ & d & $-0.2528$  (2)&$-0.2519$  (2) \\
$a_2$ & d & $-3.26  $  (3)&$-3.59  $  (4) \\
$a_3$ & d & $ 5.88  $  (6)&$ 6.5   $  (1) \\
$s_3$ & - & $-0.00463$ (2)&$-0.00459$ (2) \\
$s_4$ & - & $ 0.02164$ (3)&  0.02159  (3) \\
$ w $ & - & $-0.2282$  (f)& $-0.2282$ (f) \\
$R/n$ &-&$8.8\times10^{-17}$&$1.9\times10^{-17}$\\
$L_{\rm min}$&-& 8        &  9            \\
\hline
\end{tabular}
\end{table}
These data lead to a result for the  universal ratio $s_3/w^3\approx 0.39$,
again close to the exact \cite{Car} value $\pi/8$.

\section{Discussion}
\label{disc}
Since the corrections to scaling in $X_h(v, L)$ appear in first
order of the marginal field $v$, the data for $X_h(v, L)$ are
easier to analyze than those for the free energy, where these corrections
appear only in third and higher orders. Moreover, also the analytic
background in the free energy depends on $v \propto \Delta K^2$.
But even in the relatively simple case of $X_h(v, L)$ there are
complicating factors. There are not only contributions in different orders
of the marginal field, but also corrections due to the irrelevant fields
of the Potts fixed point. Each of these contributions displays a different
$L$-dependence. The numerical analysis has to separate all these effects,
while the range of finite sizes is quite limited. Under these conditions
a reasonably accurate and independent determination of the parameters
$u$ and $s_1$ in Eq.~(\ref{xbwfit}) was only possible because of the
accuracy of the finite-size data for $X_h(v, L)$, the knowledge
of the self-dual points of the generalized Baxter-Wu model, and the fact
that in this model the marginal field is continuously variable.
Furthermore, the independent determination of $u$ and the $s_i$
yielded universal amplitude ratios that were used in the analysis of
the four-state Potts model, ad thus enabled a determination of its
marginal field. It appears that higher-order logarithmic terms in the
expansion of $X_h(L)$ with amplitudes $s_i$ and $p_i$ are not negligible.

In the analysis of the free energy, we were not able to determine the
parameters $t_3$ and $u$ in Eq.~(\ref{yfit}) independently, but using the
result $u=-2.74$ from Sec.~\ref{numanscg}  we could still provide 
confirmations of the universal amplitude ratio \cite{Car}
$t_3/u^3=s_3/w^3=\pi/8$.

Our numerical work used real$\times$8 floating-point arithmetic and
required a computer memory of several tens of Gb. This represents what
is reasonably achievable with today's PC's. More accurate results could
be obtained in the future by using real$\times$16 floating-point arithmetic
instead, which would allow a better separation of contributions in
different orders of the marginal field.
\ack
We gratefully acknowledge the important contributions of
Prof. John Cardy to the field critical phenomena, which provided key
ingredients for much of our work.
H.~B. is indebted to Profs. Hubert Knops, Hans van Leeuwen, and Bernard
Nienhuis for enlightening comments, and acknowledges the hospitality
of the Faculty of Physics of the Beijing Normal University.
This research is supported by National Natural Science Foundation of
China under Grant No. 11175018.

\end{document}